\documentclass[a4paper,twoside]{article}
\newcommand{\affil}[1]{$^{\rm #1}$}
\usepackage{epsfig}
\usepackage[authoryear]{natbib}

\bibpunct{(}{)}{;}{a}{}{,}

\usepackage{graphicx}
\date{} 
\def\spose#1{\hbox to 0pt{#1\hss}}
\def\lta{\mathrel{\spose{\lower 3pt\hbox{$\mathchar"218$}}
     \raise 2.0pt\hbox{$\mathchar"13C$}}}
\def\gta{\mathrel{\spose{\lower 3pt\hbox{$\mathchar"218$}}
     \raise 2.0pt\hbox{$\mathchar"13E$}}}

\def\px{\phantom{0}}
\def\py{\phantom{00}}

\def\araa{{ARA\&A}}%
\def\apj{{ApJ}}%
\def\apjl{{ApJ}}%
\def\apjs{{ApJS}}%
\def\apss{{Ap\&SS}}%
\def\aap{{A\&A}}%
\def\mnras{{MNRAS}}%
\def\pasp{{PASP}}%
\title{\large\bf\flushleft
Planet-Induced Emission Enhancements in HD~179949: \\
Results from McDonald Observations}

\author{\parbox{\textwidth}{\flushleft
\vspace{-0.5cm}
{\it L. Gurdemir\affil{A}, S. Redfield\affil{B}, and M. Cuntz\affil{A,C}}\\
\vspace{0.4cm}
{\small \affil{A}\,Department of Physics, University of Texas at Arlington, Arlington,
                   TX 76019, USA} \\
{\small \affil{B}\,Astronomy Department, Van Vleck Observatory, Wesleyan University, Middletown, CT 06459, USA} \\
{\small \affil{C}\,Corresponding author. Email: cuntz@uta.edu}
}}
\begin{document}
\twocolumn[
\begin{changemargin}{.8cm}{.5cm}
\begin{minipage}{.9\textwidth}
\vspace{-1cm}
\maketitle
%
%
\small{\bf Abstract:}
We monitored the Ca~II H and K lines of HD~179949, a notable star
in the southern hemisphere, to observe and confirm previously
identified planet induced emission (PIE) as an effect
of star-planet interaction. We obtained high resolution spectra
($R \sim 53,000$) with a signal-to-noise ratio S/N $\gta 50$ in the
Ca~II H and K cores during 10 nights of observation at the McDonald
Observatory.  Wide band echelle spectra were taken using the 2.7~m
telescope.  Detailed statistical analysis of Ca~II K revealed
fluctuations in the Ca~II K core attributable to planet induced
chromospheric emission.  This result is consistent with previous
studies by \cite{shk03}.  Additionally, we were able to confirm
the reality and temporal evolution of the phase shift of the maximum
of star-planet interaction previously found.  However, no identifiable
fluctuations were detected in the Ca~II H core.
The Al~I $\lambda$3944~{\AA} line was also monitored to gauge if
the expected activity enhancements are confined to the chromospheric
layer.  Our observations revealed some variability, which is 
apparently unassociated with planet induced activity.

\medskip{\bf Keywords:}
          radiation mechanisms: nonthermal ---
          planet-star interactions ---
          stars: activity ---
          stars: chromospheres ---
          stars: individual (HD~179949) ---
          stars: late-type



\medskip
\medskip
\end{minipage}
\end{changemargin}
]
\small


\section{Introduction}

Observational results indicate that the large majority of planets outside
the solar system is hosted by main-sequence stars \citep[e.g.][]{but06}, which
are known to possess an extended outer atmosphere consisting of a chromosphere,
transition region and corona as well as a dynamic stellar wind region
\citep[e.g.][]{lin80,sim86} associated with complex magnetic structure
\citep[e.g.][]{sch00}.  In various systems, consisting of a star and a
close-in extrasolar giant planet (CEGP), interaction between the planet and
stellar atmospheric structure has been identified that is often broadly
classified as ``planet-induced (stellar) emission" (PIE).  The reality
of this phenomenon has been proposed by \cite{cun00} who also provided
a detailed ranking of the relative strength of this effect for the
various star-planet systems known at the time.  Moreover, \cite{cun00}
explored the possibility of tidal interaction and magnetic interaction
with the latter distinguished from the former by two activity maxima
per stellar circumference instead of one.

The PIE effect was subsequently discovered by \cite{shk03} in regard
to the HD~179949 system using high-resolution ($R = 110,000$) and
high signal-to-noise ratio spectra based on three observing runs at
the Canada-France-Hawaii Telescope (CFHT).  They also concluded
that the PIE effect is expected to be magnetic in nature rather than
caused by gravitational star-planet interaction.  Subsequent work was
pursued by \cite{shk05}, \cite{shk08} and others, resulting in the
identification of the PIE effect in at least six systems, which are:
HD~209458, $\upsilon$~And, $\tau$~Boo, HD~179949, HD~189733,
and HD~73526.

The latter study also provides evidence of the
on/off nature of the PIE effect, also previously found in the
$\upsilon$~And system \citep{shk08}, which according to the authors
is likely attributable to the changing stellar magnetic field
structure throughout the stellar activity cycle \citep[e.g.][]{lan10}.
An alternative or supplementary explanation of the significant
time-dependent variations of the PIE phenomena refers to the
flare-type nature of the interaction \citep[e.g.][]{saa04,mci06}.
Recently, a detailed time-dependent MHD simulation for the interaction
of the stellar magnetic field and wind with the planetary magnetosphere
and outflow was given by \cite{coh11} for HD~189733 as an example.
Their results show reconnection events occurring at specific planetary
orbital phases, causing mass loss from the planetary magnetosphere able
to generate hot spots on the stellar surface.  The simulations also
demonstrate that the system has sufficient energy for the hot spots to
be visible in Ca~II lines, which are expected to show stochastic
behavior.

There were also various attempts to identify the PIE effect in other
spectral regimes, including the radio regime, infrared and X-rays.
\cite{bas00} conducted a search for radio emission
for six extra-solar star-planet or brown dwarf systems that included
four systems with the planet or brown dwarf closer than 0.1~AU from its
host star.  No detections were made, however, which was attributed to
physical or instrumental reasons.  Subsequent work mostly focused on
the derivation of upper limits of emergent radio emission from known
systems given by, e.g. \cite{far03}, \cite{zar07}, and \cite{gri07}.

\begin{table*}
\begin{center}
\caption{Stellar and Planetary Parameters\label{tab1}}
\begin{tabular}{lcc}
\noalign{\smallskip}
\hline
\hline
\noalign{\smallskip}
Parameter              & HD~179949 & Reference \\
\noalign{\smallskip}
\hline
\noalign{\smallskip}
 Spectral Type         &  F8.5~V                                             & $^a$         \\
 RA                    &    19$^{\rm h}$ 15$^{\rm m}$ 33.2278$^{\rm s}$      & $^b$         \\
 DEC                   &  $-$ 24$^\circ$ 10$^\prime$ 45.668$^{\prime\prime}$ & $^b$         \\
 V                     &  6.25     $\pm$ ...                                 & $^b$         \\
 $T_{\rm eff}$~(K)     &  6202     $\pm$ 52                                  & \cite{rib03} \\
 $M_\ast$~($M_\odot$)  &  1.24     $\pm$ 0.10                                & \cite{tin01} \\ 
 Distance~(pc)         &  27.0     $\pm$ 0.5                                 & $^b$         \\
 Age~(Gyr)             &  2.05     $\pm$ ...                                 & \cite{don93} \\
 $R_\ast$~($R_\odot$)  &  1.193    $\pm$ 0.030                               & \cite{rib03} \\
 $[$Fe/H$]$            &  0.226    $\pm$ 0.050                               & \cite{gon07} \\
 $a_p$~(AU)            &  0.0443   $\pm$ 0.0026                              & \cite{but06} \\
 $e_p$                 &  0.022    $\pm$ 0.015                               & \cite{but06} \\
 $P_{\rm orb}$~(days)  &  3.092514 $\pm$ 0.000032                            & \cite{but06} \\
 $M_p {\sin}i$~($M_J$) &  0.916    $\pm$ 0.076                               & \cite{but06} \\
\noalign{\smallskip}
\hline
\end{tabular}
\medskip\\
$^a$data from SIMBAD; see {\tt http://simbad.u-strasbg.fr}. \\
$^b$adopted from the {\it Hipparcos} catalogue; see \cite{esa97}.
\end{center}
\end{table*}

Another effort to identify the PIE effect was undertaken by \cite{saa01}
who executed a search for periodicities in the  Ca~II infrared triplet
emission for a sample of stars with close-in giant planets; however,
no unequivocal identification was made.  Subsequently, an effort to
search for the PIE effect in the UV using the Goddard High Resolution
Spectrograph (GHRS) on board of the Hubble Space Telescope (HST) was
rendered impossible as the GHRS was removed during HST Servicing Mission 2.
Recently, \cite{gon11} reported differences regarding the chromospheric
activity between samples of stars with and without planets, although
the interpretation of these trends, if confirmed, remains unclear.

\cite{kas08} presented a statistical survey of the X-ray fluxes from
stars with close-in planets; arguably, this survey has been highly useful
as these fluxes were found enhanced by 30\% to 40\% on average over
typical fluxes from similar stars with planets that are not close-in.
On the other hand, contrary evidence was subsequently identified by
\cite{pop10}, \cite{pop11}, and \cite{can11}; nonetheless, these studies,
if found conclusive, do not exclude the possibility of enhanced X-ray
fluxes for stars with planets that are very close-in (i.e., $\lta$ 0.05~AU).
Specifically, direct X-ray observations of the HD~179949 system by
\cite{saa08} revealed a $\sim$30\% increase in the X-ray flux over
quiescent levels coincident with the phase of the Ca~II enhancements.
Additionally, the observations indicate a trend for the emission to be
hotter at increased fluxes, confirmed by modeling, showing the enhancement
at $\sim$1 keV compared to $\sim$0.4 keV for the background star.

\begin{figure}
\centering
\begin{tabular}{c}
\epsfig{file=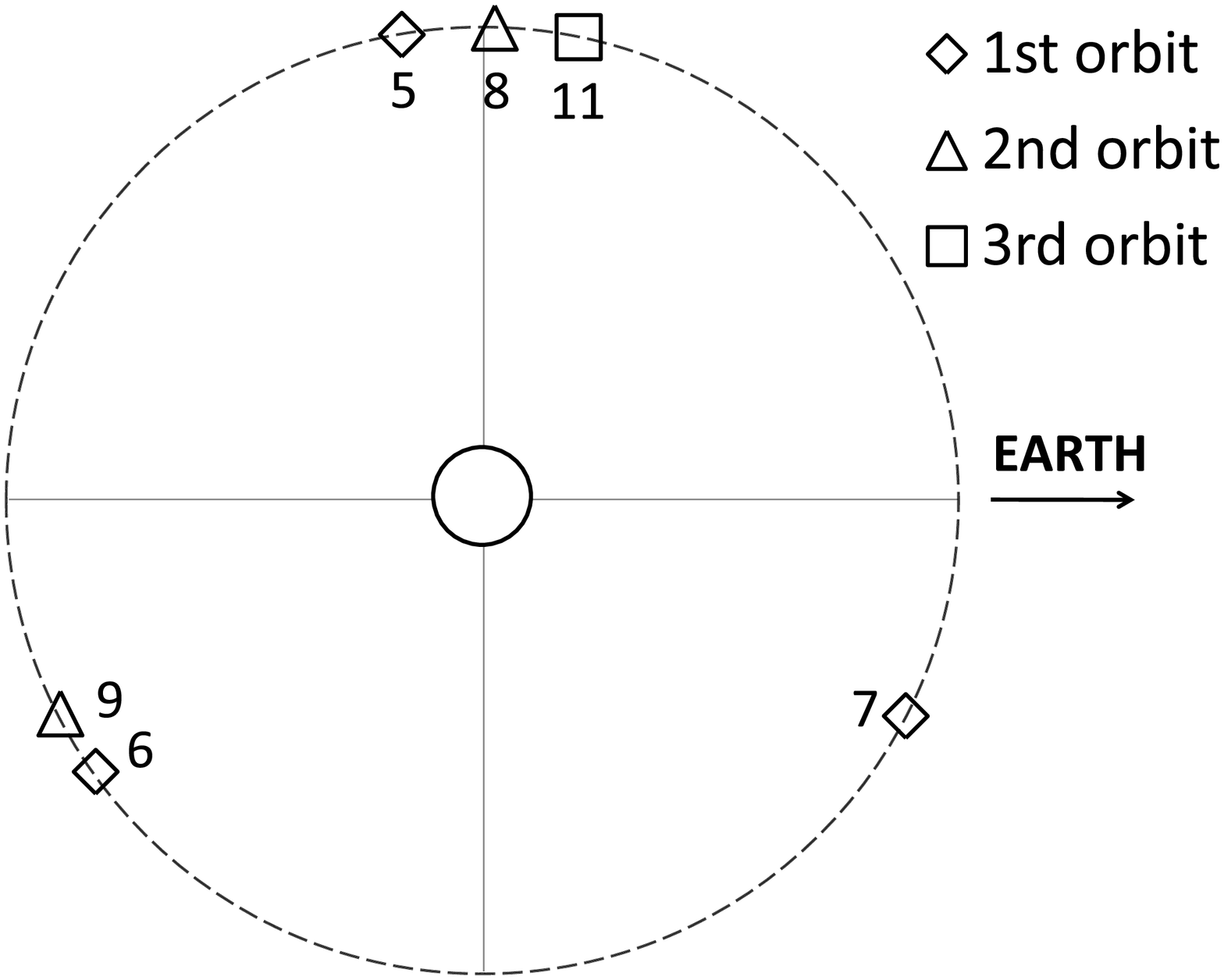,width=0.85\linewidth}
\end{tabular}
\caption{
Orbital position of the planet around its host star (HD~179949) during
the observational run.  Note that $\phi = 0$ corresponds to the planetary
position between the host star and Earth (sub-planetary point).
The numbers indicate the dates in April 2006 (local time) when the
observations were obtained.
}
\end{figure}

In our present study, we will again focus on the HD~179949 system.
HD~179949 is an F8.5~V star at a distance of $27.0 \pm 0.5$~pc
located in the southern hemisphere in the constellation Sagittarius
with a magnitude of V = 6.25.  Its mass is estimated as
1.24 $M_\odot$ \citep{tin01} based on the interpolation of
evolutionary tracks previously computed by \cite{fuh97,fuh98}.
The age of HD~179949 has been estimated as 2.05~Gyr
(see Table~1) based on the strength of the chromospheric
component of the Ca~II emission, $R'_{\rm HK}$ \citep{noy84,don93}.
Note that HD~179949 is a solar-type star of moderate age, implying
that its inherent chromospheric emission and variability has
considerably subsided\footnote{See \cite{gui03} and \cite{gue07},
among other contributions to the ``Sun in Time" project, for
studies on the decline of chromospheric emission on stellar
evolutionary time scales.}, thus adding to the credibility
of any PIE determinations.  By employing the Doppler velocity
technique,  \cite{tin01} discovered as part of the
Anglo-Australian Planet Search a ``Hot Jupiter" hosted by
HD~179949 located at a distance of about 0.044~AU (7.99 stellar radii).
The planetary orbit has a very low eccentricity that is essentially
consistent with the circular case (see Table~1), allowing us to
ignore the orbital ellipticity, if any, in the following.
The planetary mass parameter is given as $M_p~{\sin}i =
0.916 \pm 0.076$ $M_J$.

\begin{table*}
\begin{center}
\caption{Details of Observations\label{tab2}}
\begin{tabular}{lccccc}
\noalign{\smallskip}
\hline
\hline
\noalign{\smallskip}
Number        &
Date          &
Time          &
Exposure Time &
S/N           &
Seeing        \\
...           &
...           &
(UT)          &
(s)           &
...           &
(arcs)        \\
\noalign{\smallskip}
\hline
\noalign{\smallskip}
  1  &   April 5, 2006  &  11:31  &  1200	&  53	 &  2.6  \\
  2  &   April 5, 2006	&  11:54  &	 1200 &  56	 &  2.6  \\
  3  &   April 6, 2006	&  11:17  &  1200	&  43  &  4.1  \\
  4  &   April 6, 2006	&  11:40  &  1200	&  47	 &  4.1  \\
  5  &   April 7, 2006	&  11:30  &  1200	&  37	 &  5.0  \\
  6  &   April 7, 2006	&  11:53  &  1200	&  45	 &  5.0  \\
  7  &   April 8, 2006	&  11:13  &  1200	&  50	 &  2.3  \\
  8  &   April 8, 2006	&  11:36  &  1200	&  51	 &  2.3  \\
  9  &   April 8, 2006	&  11:59  &  1200	&  70	 &  2.3  \\
 10  &   April 9, 2006	&  11:59  &  1200	&  44	 &  2.1  \\
 11  &  April 11, 2006	&  11:31  &  1200	&  51	 &  1.7  \\
 12  &  April 11, 2006	&  11:54  &  1200	&  66	 &  1.7  \\
\noalign{\smallskip}
\hline
\end{tabular}
\end{center}
\end{table*}

\begin{table*}
\begin{center}
\caption{Ephemerides of HD~179949b\label{tab3}}
\begin{tabular}{lccc}
\noalign{\smallskip}
\hline
\hline
\noalign{\smallskip}
Date            &
JD              &
Orbital Period  &
Reference       \\
...             &
...             &
(d)             &
...             \\
\noalign{\smallskip}
\hline
\noalign{\smallskip}
   07 / 1998  &  2451001.510 	&  3.092514     &  \cite{but06}$^a$ \\  			
   09 / 2000  &  2451793.980 	&  3.0925{\py}  &  $^b$... \\
   07 / 2002  &  2452479.823 	&  3.09285{\px} &  \cite{shk03} \\
   09 / 2003  &  2452894.110 	&  3.09246{\px} &  \cite{shk03} \\
\noalign{\smallskip}
\hline
\end{tabular}
\medskip\\
$^a$based on data obtained by \cite{tin01}. \\
$^b${\tt http://www.exoplanet.eu}.
\end{center}
\end{table*}

\begin{figure}
\centering
\begin{tabular}{c}
\epsfig{file=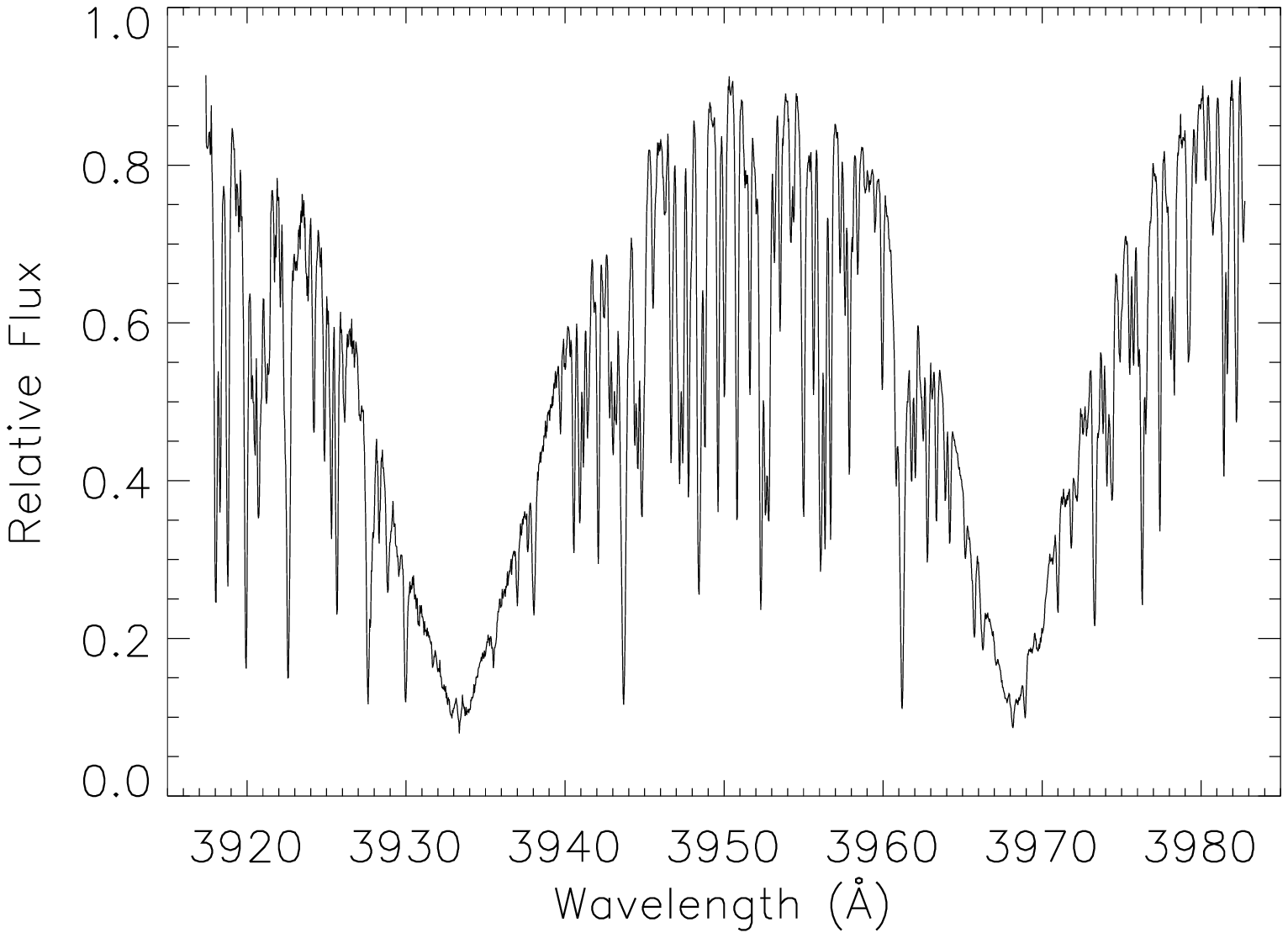,width=0.95\linewidth} \\
\epsfig{file=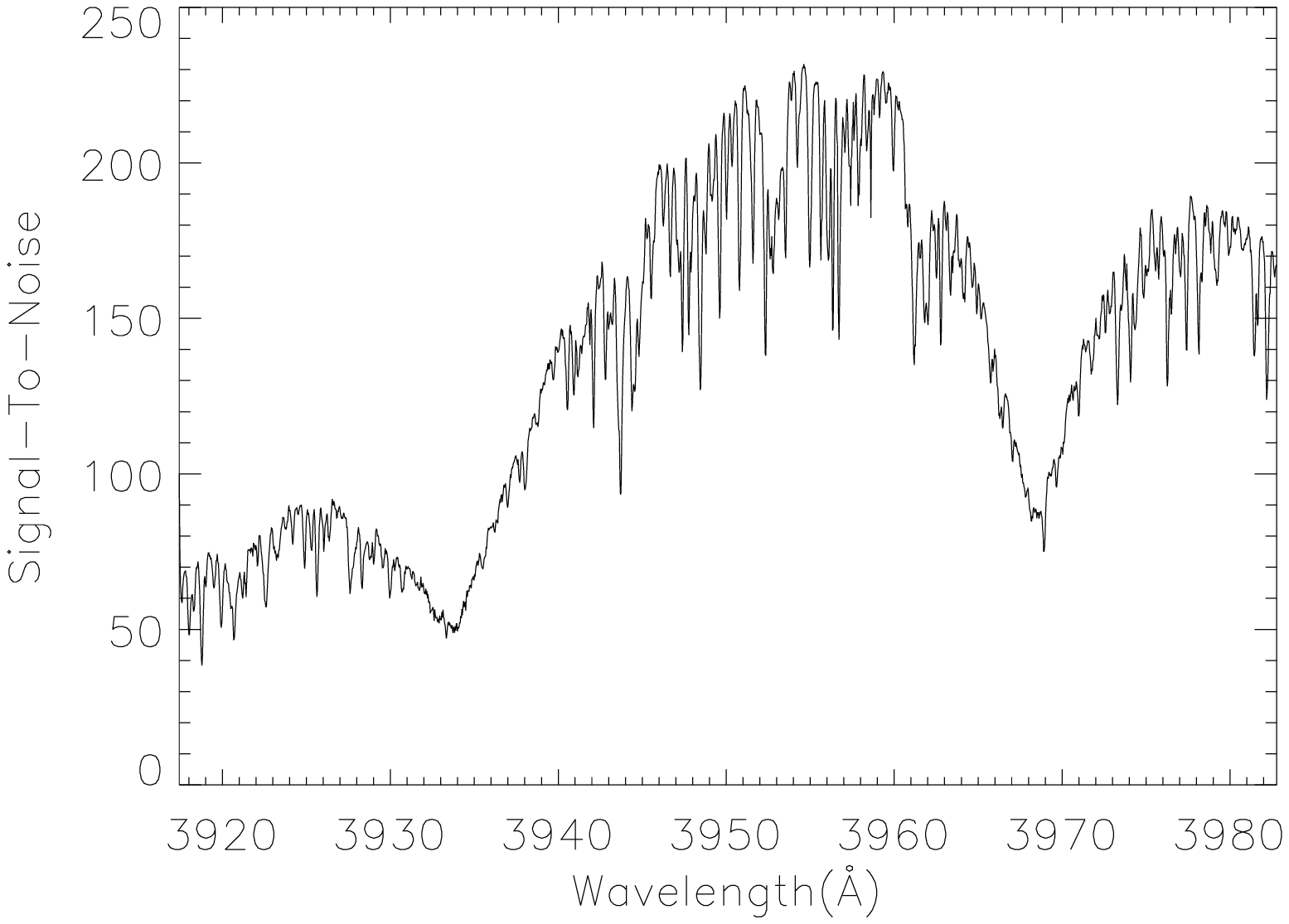,width=0.95\linewidth}
\end{tabular}
\caption{
Single flat-fielded and blazed spectrum of HD~179949 taken on
April 5, 2006 with an exposure time of 1200~s.  We depict the
Ca~II regime (3915 -- 3985 {\AA}) of the wide band echelle spectrum
({\it top}) and the associated S/N ratios ({\it bottom}).
}
\end{figure}

Significant evidence that the activity enhancement in HD~179949 is
planet-induced is based on the finding that the stellar rotation period
$P_{\rm rot}$ almost certainly disagrees with the orbital period of the
planet.  Therefore, it would be difficult to explain emission on the
star that essentially ``follows" the CEGP without a planet-related cause.
Unfortunately, HD 179949 does not have a robustly measured $P_{\rm rot}$ \citep[see][]{wol04}.
However, following \cite{saa04} and references therein, it can be
argued based on the strength of the Ca~II HK emission, the mean
Mount Wilson index $<S>$ as well as stellar evolution constraints
that $P_{\rm rot}$ lies between 7 and 10~d, thus being distinctly
different from $P_{\rm orb} \simeq 3.092$~d (see Table~1).

The PIE effect has various interesting astrophysical applications as
it, in principle, may allow the identification and characterization
of magnetic fields and dynamo activity in exosolar Jupiter-type planets
\citep[e.g.][]{ip04,gri04,erk05,zar07,lan08,coh09,sch10,lan11,coh11}.
In addition, it may also be useful for probing the inner region of
the stellar winds \citep[e.g.][]{pre05,mci06}.
This is a further reason to continue monitoring the PIE effects in
different spectral regimes and by using different type of instruments
including the 2.7~m Harlan J. Smith Telescope at the McDonald Observatory.
The latter allowed us to continue studying the Ca~II H and K line cores
in HD~179949 and to compare our findings with the previous results by
Shkolink et al. (2003) and subsequent work.  Our paper is structured
as follows:  In \S 2, we describe the employed methods including the
data reduction.  Our results and discussion are given in \S 3.  In \S4,
we present our conclusions.


\section{Methods and Data Reduction}

In order to study the PIE effect in the HD~179949 system, we obtained
high resolution ($R \equiv \lambda/\Delta\lambda \sim $53,000) spectra using the McDonald Observatory's echelle
spectrograph mounted on the TK3 light path of the Coud\'e focus of
the 2.7~m (107~inch) Harlan J. Smith Telescope.  We obtained
observations for six (almost) consecutive nights in April 2006, which allowed
us to acquire data points spreading over nearly three planetary orbits
(see Fig.~1 and Table~2 for details).
Full optical bandwidth (3400 --– 10,900 {\AA}) were recorded
in at least 67 orders by a Tektronix (2048 $\times$ 2048),
CCD camera.  The Ca~II H and K lines were carefully centered
in order 56.
Furthermore, Th-Ar spectra were taken frequently (i.e., typically
in less than 2~h intervals) to allow accurate wavelength calibrations.
The signal-to-noise (S/N) ratio of the spectra exceeded 200 for the
continuum, and it was close to 50 in the Ca~II~K core (3933 {\AA}),
and about 80 in the Ca~II~H core (3968 {\AA}) (see Fig.~2).

The data reduction was performed by using the IRAF standard packages
IMRED, CCDRED and ECHELLE.  The data reduction procedures can be
summarized as follows: Combined biases were subtracted from all stellar,
arc, and flat-field exposures to remove the baseline noise.  Scattered
light was removed from all stellar and flat field exposures.
All stellar and arc exposures were divided by combined flat field for
normalization of the sensitivity levels of the CCD chip.  Spectra were
extracted from the 2D images to 2D spectra.
Heliocentric correction was applied to 
each spectrum to express the observation time considering that the
observer's location does not coincide with the center of the Sun.
Differential radial velocity corrections due to the rotation
and orbital velocity of Earth were applied to 
each spectrum.  The combined flat field was processed like a
stellar image to obtain blaze functions for each observation night.
Blaze functions were used to remove from each final spectrum the
low-order curvature as a result of the intensity variation along
the orders.

\begin{figure}
\centering
\begin{tabular}{c}
\epsfig{file=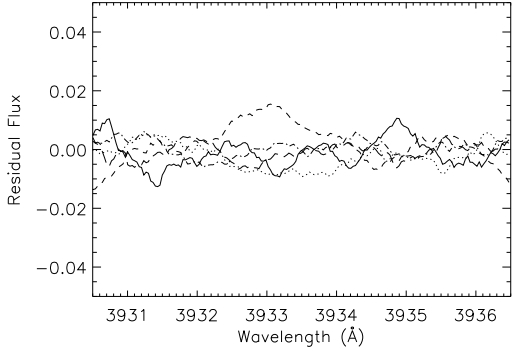,width=0.95\linewidth}
\end{tabular}
\caption{
Residual flux of the Ca~II~K core of HD~179949 between April 5, 2006 and
April 11, 2006.  The residual flux is computed from the average mean and
smoothed over 21 pixels.  The results are indicated by solid (April 5),
dotted (April 6), dashed (April 7), dash-dotted (April 8), and long-dashed
(April 11) lines.  Note that the data of April 9 have been disregarded.
}
\end{figure}

\begin{figure}
\centering
\begin{tabular}{c}
\epsfig{file=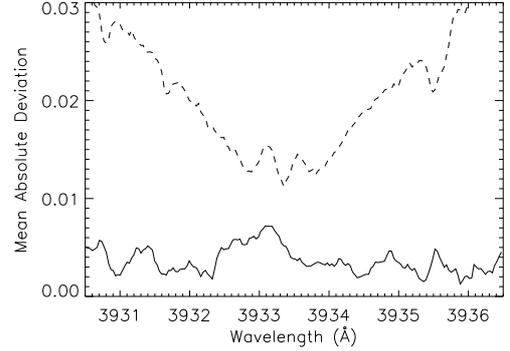,width=0.95\linewidth}
\end{tabular}
\caption{
The solid line indicates the Mean Absolute Deviation (MAD) of the
Ca~II~K core of HD~179949 as observed.  The MAD values are calculated as
described by \cite{shk03}, noting that
${\rm MAD} = N^{-1} \sum_{i=1}^N {\vert data_i - mean \vert}$ based on
$N$ spectra.  The unit is the relative intensity attained as a fraction
of the normalized flux.  The dashed line represents the mean spectrum of
the Ca~II~K core (scaled accordingly) to show that the activity is confined, slightly asymmetrically, 
to the Ca~II~K core.  The data of April 9 have been disregarded.
}
\end{figure}

Each reduced spectrum is carefully checked for potential contaminations.
Trace amounts of sunlight contamination are detected in the last exposures
of each observing night as HD~179949 is observed at low altitudes on the
Eastern horizon just a few hours before dawn.  These exposures were
discarded to avoid any solar contamination of our spectra.
Although the night of April 9 originally indicated a critical phase to
detect star-planet interaction, this entire data set had to be discarded
as the cloud coverage caused large noise level in the spectra.

Next we followed the procedure of \cite{shk03} previously used to unveil
the PIE effect for the HD~179949 system from optical spectra.
The following steps were undertaken: (1) From each spectrum increments
of 7~{\AA} centered at the Ca~II~H (3968 {\AA}) and K (3933 {\AA}) cores,
respectively, were extracted. (2) To normalize each spectrum, the
end points of each extracted spectra were set to 1.  As stated by
\cite{shk03}, this is the best way to normalize the spectra in the
wavelength regimes, such as H and K lines,
where there is no clear continuum. (3) The
normalized spectra were grouped by date and co-added in order to
obtain a mean spectrum for each night.  (4) Overall average spectra
were computed over all nights.  (5) Residuals given as
$(data_i - mean)$ were computed.  (6) Low-order curvature was
removed from each residual spectra.  This step was performed by
applying fits to the data sets with polynomials of order 2.
In order to obtain the best trend line for the low-order curvature,
data points in the proximity of the center line were excluded from the
fit.  (7) Next the residual spectra is smoothed over 21 pixels.
(8) Finally, the Mean Absolute Deviation is computed according to
${\rm MAD} = N^{-1} \sum_{i=1}^N {\vert data_i - mean \vert}$ with
$N$ denoting the number of spectra.  As a further part of the process,
all spectra were analyzed with respect to 7~{\AA} window.  Trapezoidal
integration was used to compute the integrated K residuals.  Error bars 
were calculated from nightly variations.


\begin{figure}
\centering
\begin{tabular}{c}
\epsfig{file=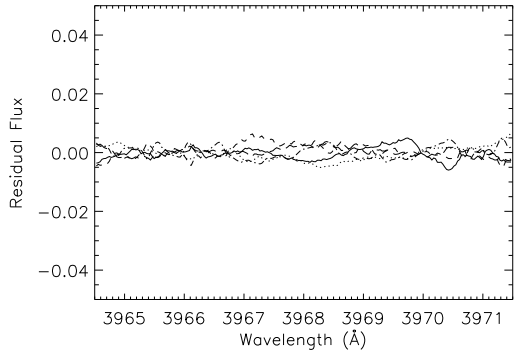,width=0.95\linewidth}
\end{tabular}
\caption{
Same as Fig.~3, but now for the Ca~II~H core of HD~179949.
}
\end{figure}

\begin{figure}
\centering
\begin{tabular}{c}
\epsfig{file=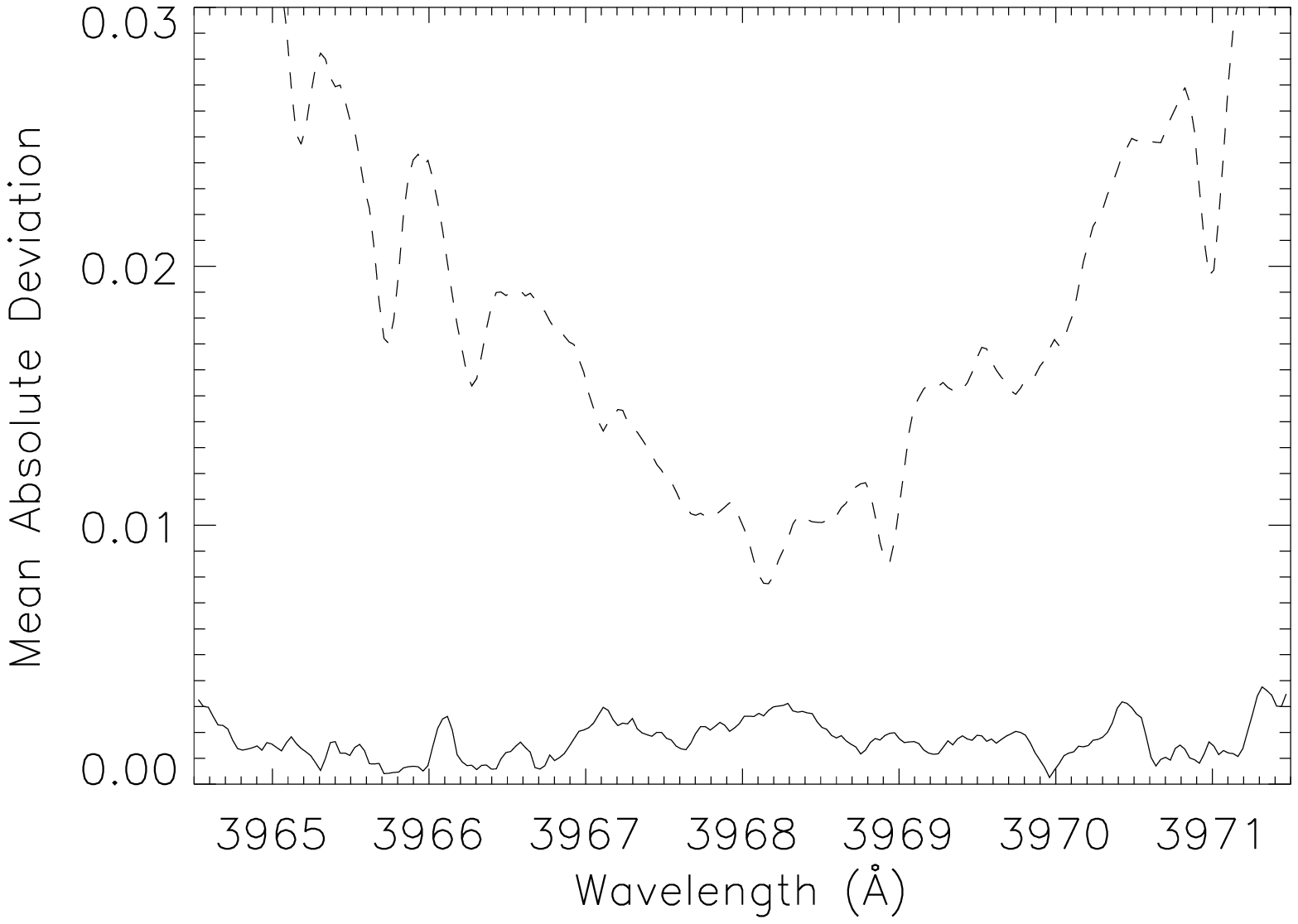,width=0.95\linewidth}
\end{tabular}
\caption{
Same as Fig.~4, but now for the Ca~II~H core of HD~179949.
}
\end{figure}

\begin{figure}
\centering
\begin{tabular}{c}
\epsfig{file=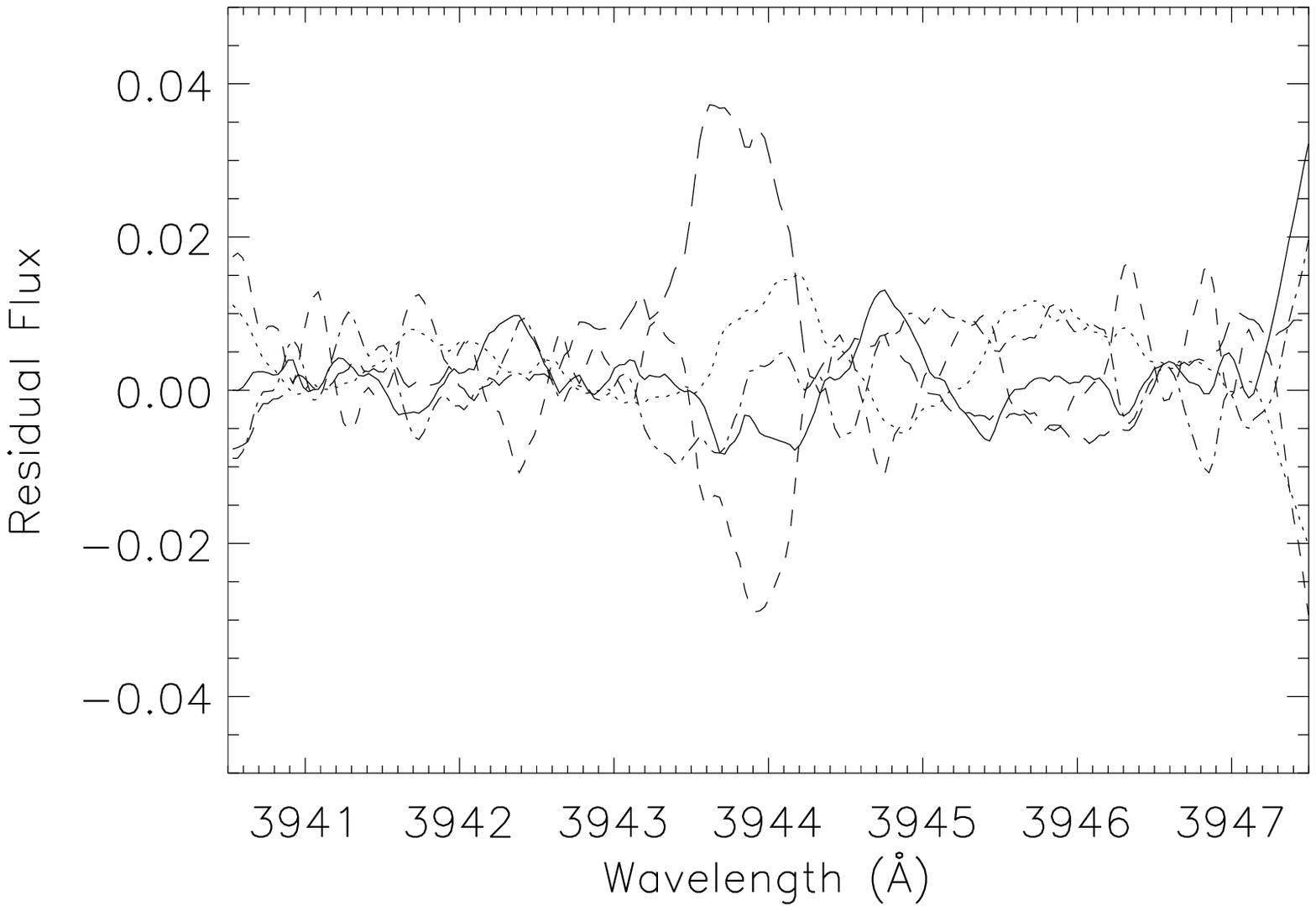,width=0.95\linewidth} \\
\epsfig{file=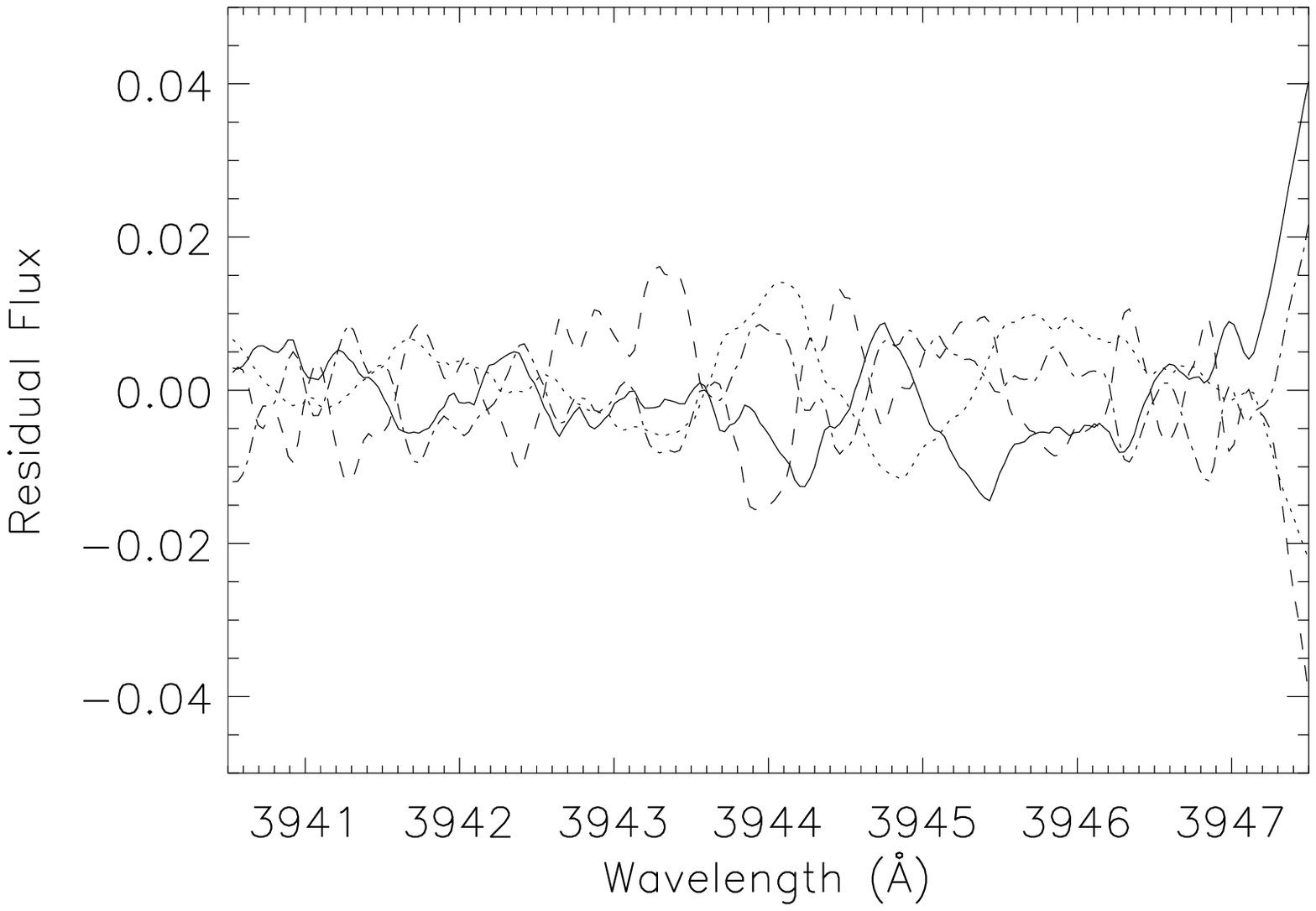,width=0.95\linewidth}
\end{tabular}
\caption{
{\it Top:} Residual flux of the Al~I $\lambda$3944~{\AA} line between
April 5, 2006 and April 11, 2006 akin to Fig. 3 noting that the data of
April 9 have again been disregarded.  {\it Bottom:} Here we also
disregarded the data of April 11, noting these data points either suffer from an unknown data artifact or are 
due to photospheric activity unrelated to star-planet
interaction (see text).  This signal has no counterpart in 
Ca~II H and K.
}
\end{figure}

\section{Results and Discussion}

The nightly flux residuals in the K core are over-plotted in Fig.~3.
The figure indicates fluctuations in the K core apparently due
to star-planet interaction akin to the results previously obtained by
\cite{shk03}.  The calculated Mean Absolute Deviation (MAD) is shown
in Fig.~4.  The average of all normalized nightly spectra (mean values)
is depicted using dashed-lines indicating the MAD
is confined, slightly asymmetrically, to the K core.  A similar analysis has been pursued for the
Ca~II~H line at 3968~{\AA}.   The Ca~II~H line does not show any
identifiable variation (see Figs. 5 and 6).
Previously, Ca~II~H fluctuations in the line core were deduced to be
about 2/3 of the strength of the Ca~II~K core as stated by \cite{shk03};
see, e.g. \cite{lin70} for background information on Ca~II.  We currently have no explanation for this discrepancy between the predicted and observed Ca~II~H emission.

Another focus of our study is to investigate the behavior of the
Al~I $\lambda$3944~{\AA} line (see Figs. 7 to 9), which is formed in
the stellar photosphere.  If the fluctuations in Ca~II H, K are caused
by star-planet interaction, as expected, rather than by an inherently
stellar process, a significantly reduced level of variability in the
stellar photosphere is expected to occur compared to the stellar
chromosphere.  The latter is a natural consequence of the strong density
gradient between the stellar photosphere and chromosphere; see e.g.
the VAL-C model for the Sun (G2~V) \cite{ver81}, which has a similar,
albeit moderately lower surface temperature than HD~179949.  Noting
that Al~I $\lambda$3944~{\AA} is a photospheric line positioned between
the Ca~II H and K lines in the spectrum, it can thus be utilized for
distinguishing between planet-related chromospheric activity and
and planet-unrelated photospheric activity.

The Al~I line analysis is performed to isolate planet-induced
chromospheric activity from possible photospheric fluctuations.
Since the photospheric response is expected to be significantly less than the chromospheric signal, this is also a check on systematic errors in our data analysis.  Previously, \cite{shk03} stated that no variation has been detected
in 2~{\AA} window of Al~I analysis.  We extend our analysis to a
7~{\AA} window to create results comparable to the Ca~II H and K cores.
It is found that the Al~I line analysis in general does not indicate
enhanced photospheric activity, even if enhanced chromospheric activity
occurs.  A notable exception, however, is the observation on April 11, 2006,
where a strong Al~I line variation is found without a chromospheric
counterpart.  Thus, if it is indicative of a real photospheric signal, it is apparently unassociated with star-planet
interaction.

Detected photospheric emission raises the level of the mean spectrum
around the line core resulting in a greater residual flux (see Fig. 7).
As noted in Sect. 2, the residual flux is computed from $(data_i - mean)$;
therefore, the fluctuation increases because of an increased mean value.
Integrated residuals between 3943.5 {\AA} and 3944.5 {\AA} are plotted as
a function of the planetary phase (see Fig. 8).  The April 11 data,
indicated by a square point, show no correlation with the activity
enhancement detected in the K core of Ca~II.  If we ignore the April 11 data point regardless of the cause of the variability, and repeat our analysis, no significant fluctuation and/or activity enhancement is revealed
(see Fig. 7).  The Mean Absolute Deviation (MAD) is subsequently also
computed for the Al~I line (Fig. 8) with no indication of activity
enhancement found.

\begin{figure}
\centering
\begin{tabular}{c}
\epsfig{file=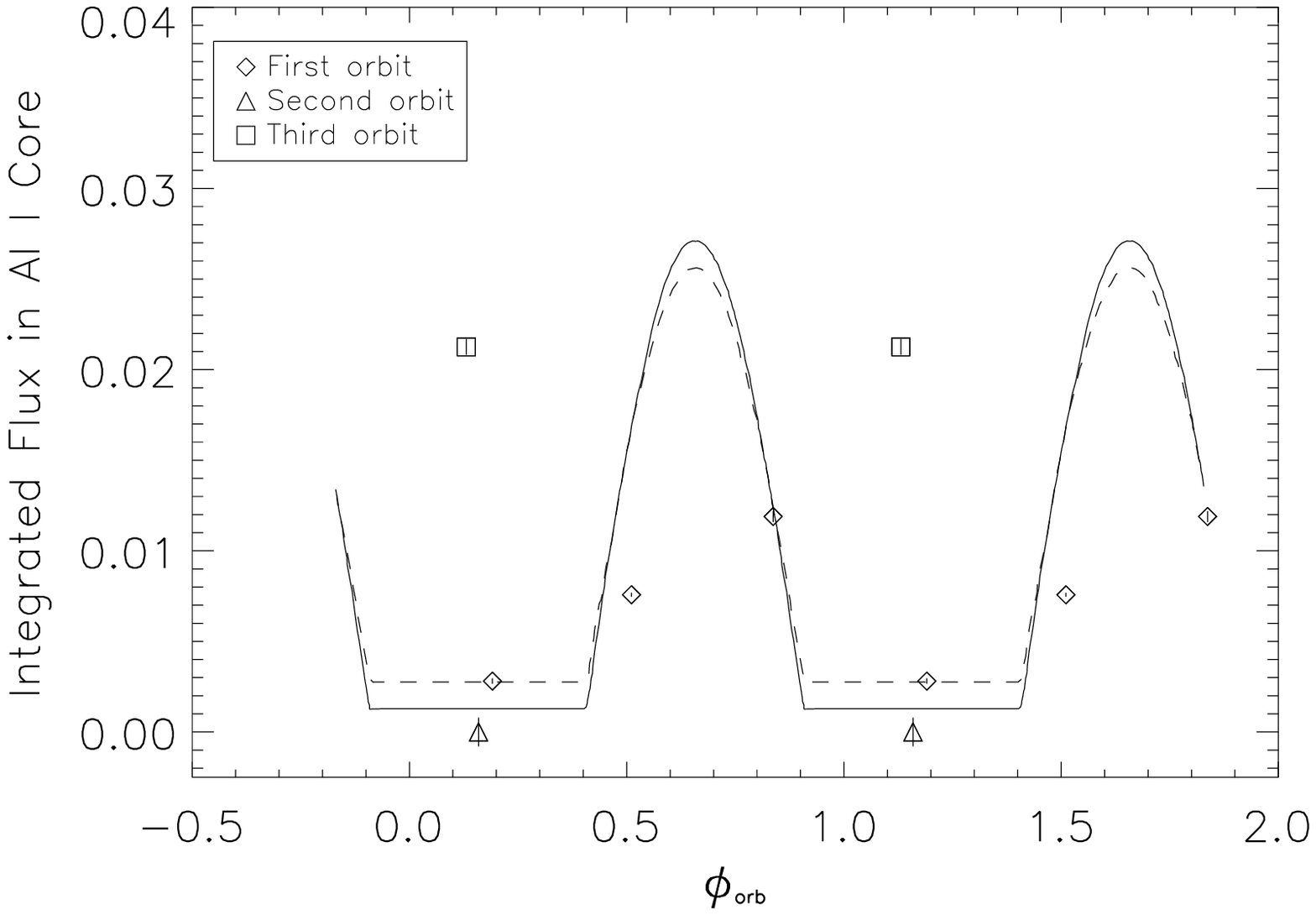,width=0.95\linewidth} \\
\epsfig{file=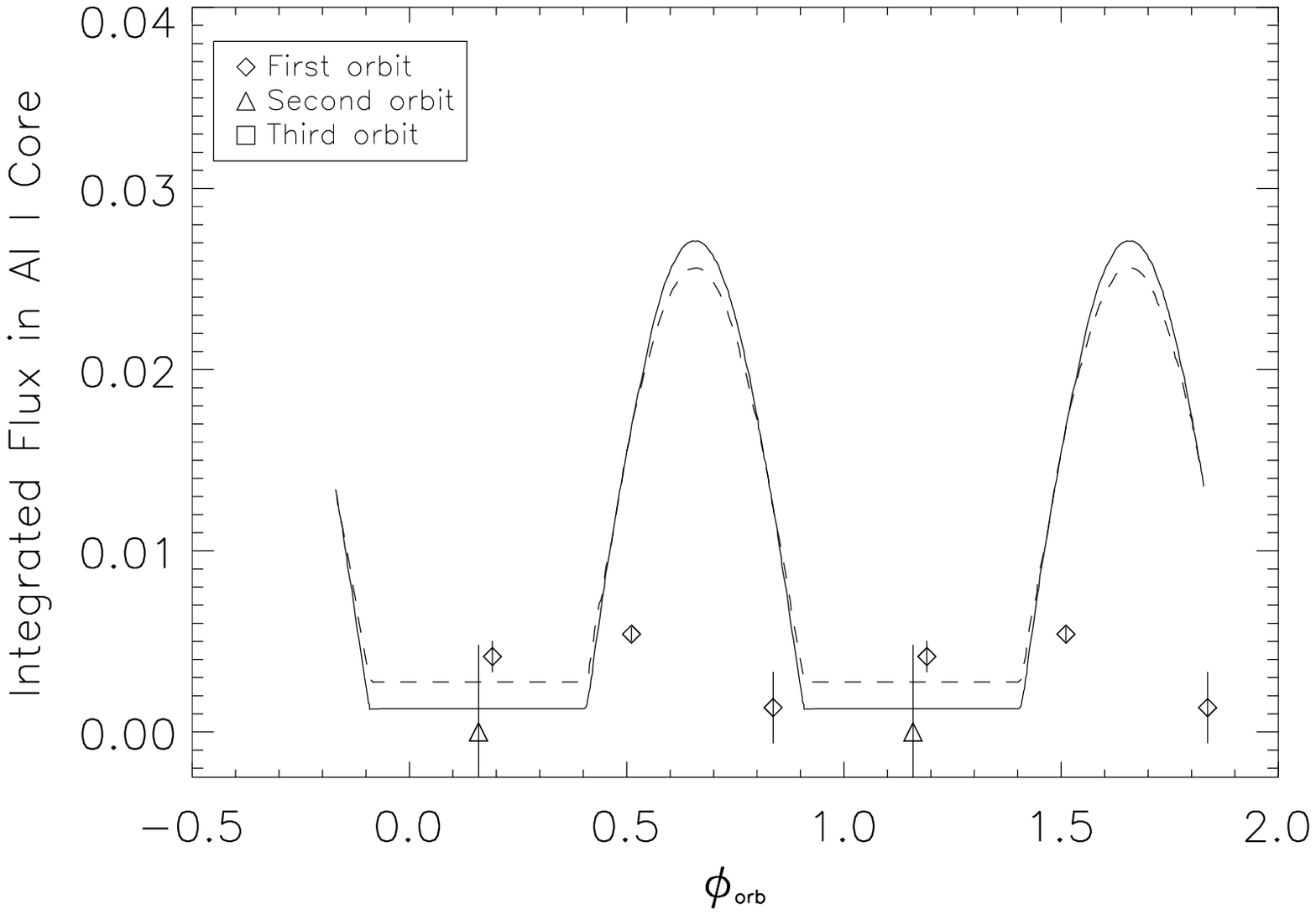,width=0.95\linewidth}
\end{tabular}
\caption{
Depiction of the integrated residual flux of the Al~I $\lambda$3944~{\AA} line
of HD~179949.  In the top figure, the data of April 9 have been disregarded,
whereas in the bottom figure the data of both April 9 and April 11 have been
omitted.  Note that the data of April 11 are most likely due to photospheric
activity unrelated to star-planet interaction.
}
\end{figure}

\begin{figure}
\centering
\begin{tabular}{c}
\epsfig{file=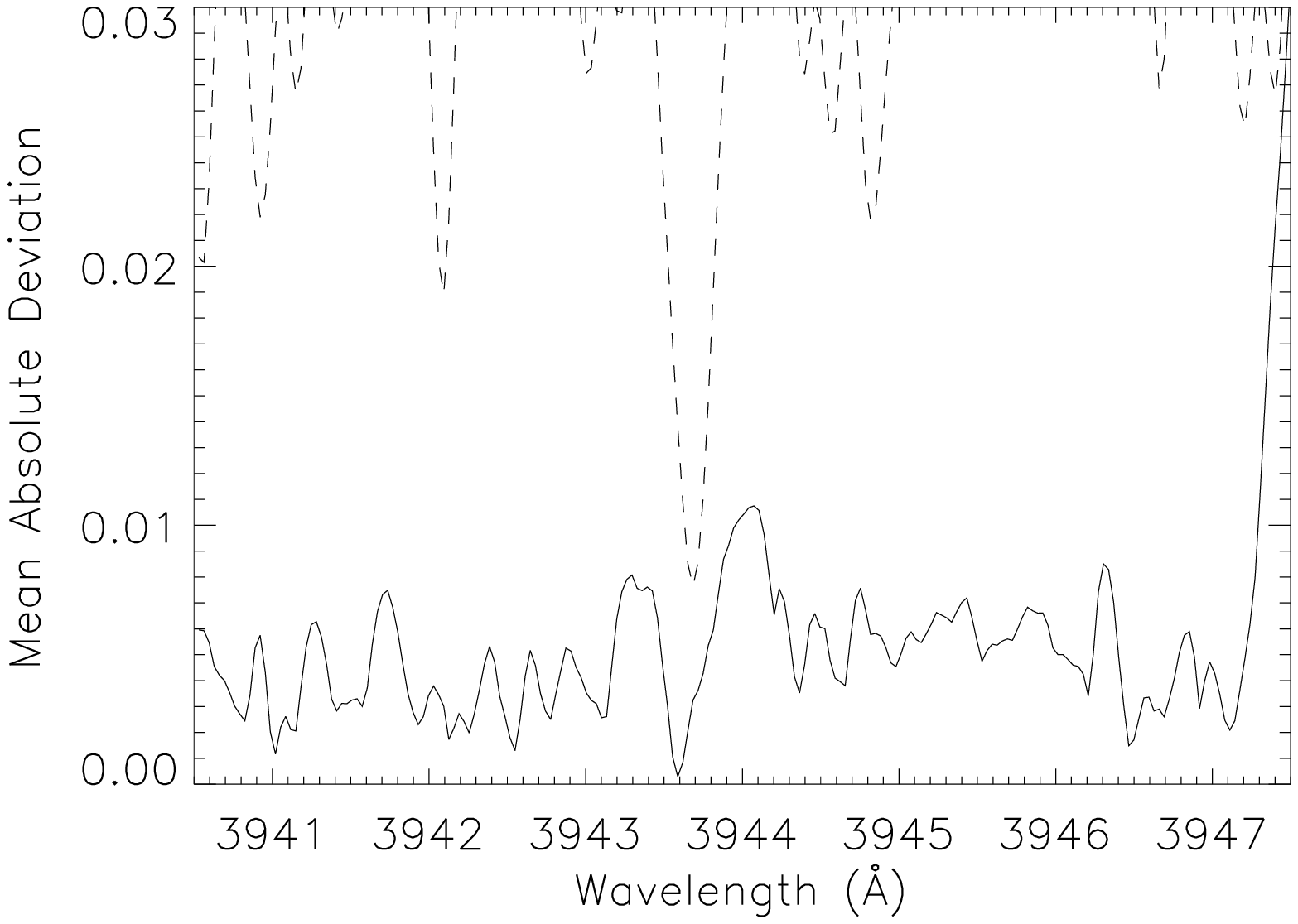,width=0.95\linewidth}
\end{tabular}
\caption{
Same as Fig.~4, but now for the Al~I $\lambda$3944~{\AA} line
of HD~179949.  Note that the data of both April 9 and April 11
have been disregarded (see text for details).
}
\end{figure}

\begin{table*}[ht]
\begin{center}
\caption{Summary of Observations$^a$\label{tab4}}
\begin{tabular}{lccccccc}
\noalign{\smallskip}
\hline
\hline
\noalign{\smallskip}
Date       &
UT         &
JD         &
HJD        &
$\phi (1)$ &
$\phi (2)$ &
$\phi (3)$ &
$\phi (4)$ \\
...        &
(h)        &
...        &
...        &
...        &
...        &
...        &
...        \\
\noalign{\smallskip}
\hline
\noalign{\smallskip}
   April 6, 2006 &  11:31 &	2453831.980  &	 2453831.980  & 0.2650 &	0.0137 &	0.1880 &	0.2750 \\
   April 6, 2006 &  11:54 &	2453831.996  &	 2453831.996  & 0.2702 &	0.0189 &	0.1931 &	0.2802 \\
   April 7, 2006 &  11:17 &	2453832.970  &	 2453832.970  & 0.5852 &	0.3339 &	0.5082 &	0.5952 \\
   April 7, 2006 &  11:40 &	2453832.986  &	 2453832.986  & 0.5904 &	0.3391 &	0.5133 &	0.6004 \\
   April 8, 2006 &  11:30 &	2453833.979  &	 2453833.979  & 0.9115 &	0.6602 &	0.8345 &	0.9216 \\
   April 8, 2006 &  11:53 &	2453833.995  &	 2453833.995  & 0.9167 &	0.6654 &	0.8396 &	0.9267 \\
   April 9, 2006 &  11:13 &	2453834.967  &	 2453834.968  & 0.2311 &	0.9798 &	0.1540 &	0.2411 \\
   April 9, 2006 &  11:36 &	2453834.983  &	 2453834.983  & 0.2362 &	0.9849 &	0.1591 &	0.2462 \\
   April 9, 2006 &  11:59 &	2453834.999  &	 2453835.000  & 0.2414 &	0.9901 &	0.1643 &	0.2515 \\
  April 10, 2006 &  11:59 &	2453835.999  &	 2453836.000  & 0.5647 &	0.3135 &	0.4877 &	0.5749 \\
  April 12, 2006 &  11:31 &	2453837.980  &     2453837.980  & 0.2052 &	0.9540 &	0.1281 &	0.2154 \\
  April 12, 2006 &  11:54 &	2453837.996  &	 2453837.996  & 0.2103 &	0.9592 &	0.1333 &	0.2206 \\
\noalign{\smallskip}
\hline
\end{tabular}
\medskip\\
$^a$$\phi (1)$, $\phi (2)$, $\phi (3)$, and $\phi (4)$ correspond to the observations obtained on
07/1998, 09/2000, 07/2002, and 09/2003, respectively, (see Table~3) with $\phi (1) \equiv
\phi_{\rm Butler}$.
\end{center}
\end{table*}

To investigate the modulation of the PIE effect with the orbital period
of the planet $P_{\rm orb}$, the Ca~II~K residuals are integrated and
plotted as a function of orbital phase (see Fig. 10).
The integrated residuals are grouped by the orbital period of
the planet.  Each orbital coverage is distinguished by different
symbols.  The scale of the figure is adjusted compared to
\cite{shk03} by setting the minimal residual flux to zero.
The best-fit bright-spot models from \cite{shk03}
are also depicted; they implement models of photospheric spatial
inhomogeneities (spaced akin to $P_{\rm rot}$) in comparison to
planetary effects (spaced akin to $P_{\rm orb}$).  The planetary
positions for our observation run, calculated from the ephemerides
of 2002 provided by \cite{shk03}, are depicted in Fig.~1.

Figure 10 indicates enhanced activity during phases 0.5 and 0.85
probably due to star-planet interaction.  This is consistent with
previous results of \cite{shk05,shk08} as the data points are
observed near bright-spot model fits.  \cite{shk05,shk08} suggested
that the peak emission is found at a phase of about $\phi = 0.8$
while stating that the fit to the 2001 and 2002 data peaks at
$\phi = 0.83 \pm 0.04$ with an amplitude of 0.027.
The phase shifts in the HD 179949 system were first detected in
previous work by E. Shkolnik and collaborators, when the observational
data spanning 6 years were analyzed.  New fit functions suggested phase
shifts of $- 0.07$ between 2002 and 2005 data points, and $- 0.17$
between 2003 and 2006 data points (Fig.~11).

\begin{figure}
\centering
\begin{tabular}{c}
\epsfig{file=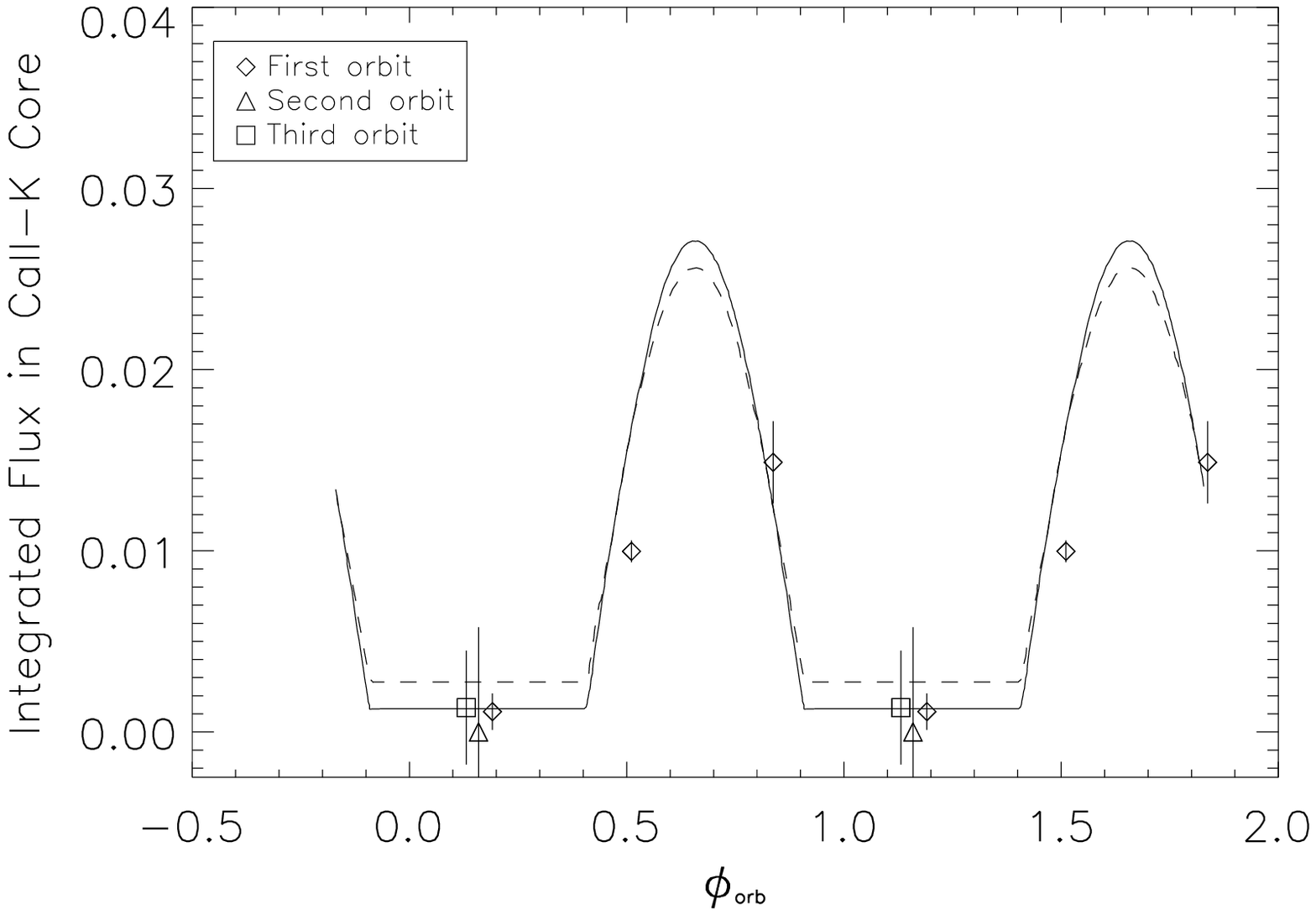,width=0.95\linewidth} \\
\epsfig{file=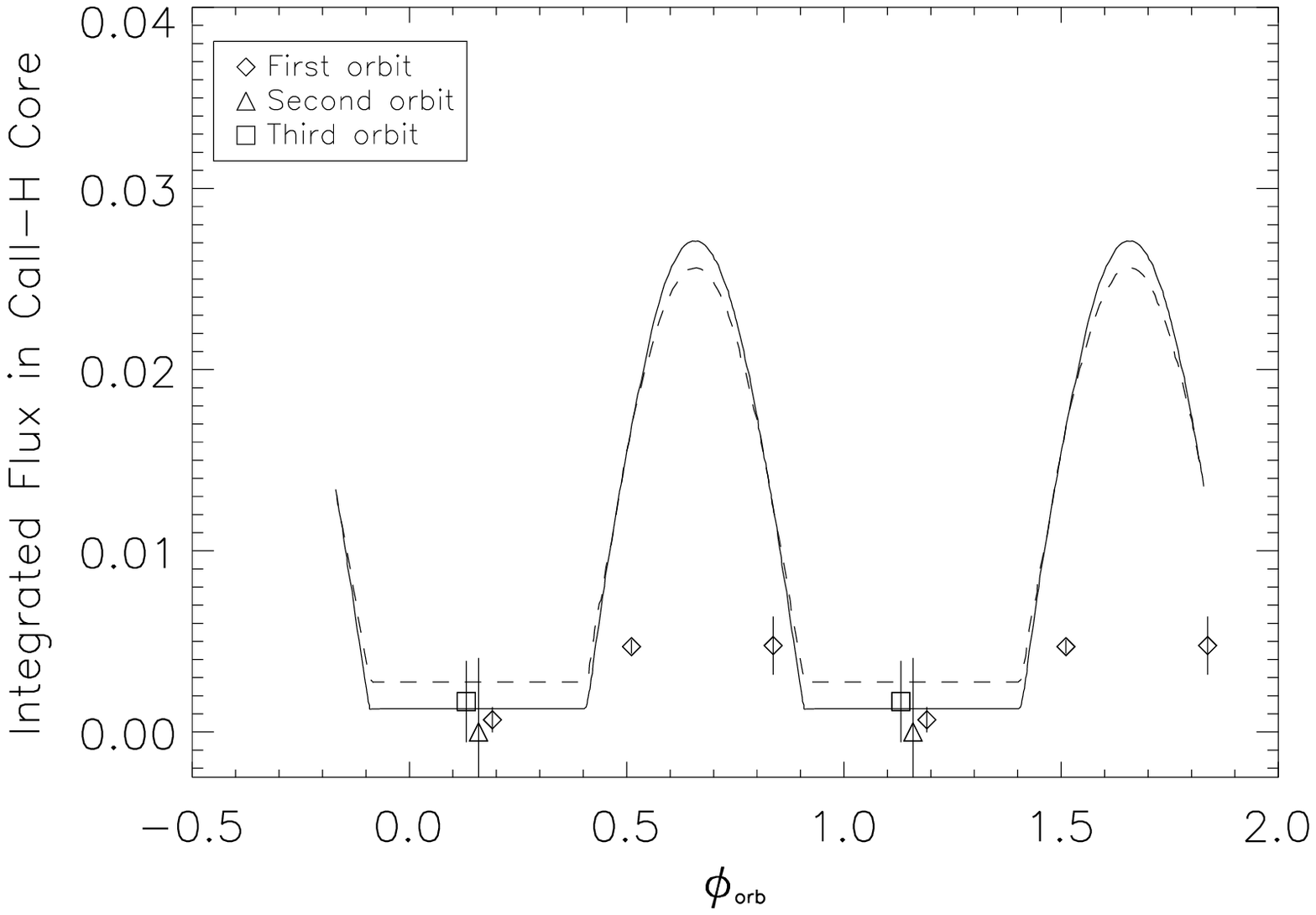,width=0.95\linewidth}
\end{tabular}
\caption{
{\it Top:}
Integrated residual flux of the Ca~II K cores as a function of the orbital
phase.  Different symbols are used to distinguish the data points from the
three consecutive orbits of HD 179949b.  All flux points are shifted
vertically to be consistent with a minimum flux of zero.  The ephemerides
for calculating the phases are those given by \cite{shk03}.  The solid line
indicates the best-fit bright-spot model as discussed by \cite{shk03} with
a spot at a latitude of 30$^{\circ}$ and a stellar inclination of
$i = 87^{\circ}$.  The dashed line represents the same model but for
$i = 83^{\circ}$.  Note that the phase of the best-fit bright-spot models
are shifted by $- 0.17$, an amount previously suggested by \cite{shk08}
in the view of activity enhancement observations of 2003 and 2006.
{\it Bottom:}
Same as top figure, but now for Ca~II H.
}
\end{figure}

Currently, the cause(s) of these phase shifts are unknown, while
noting that it may be possibly related to the star-planet interaction
itself.  As previously pointed out \citep[e.g.][]{mci06,shk08,coh09},
the planet-induced activity leads the planet, and the amount of
phase lead may permit to decipher the magnetic field geometry,
including the existence of a Parker-type spiral.  However, considering
that in HD~179949 the Parker spiral is relatively close to the star,
a potentially more likely source of activity is
magnetic reconnection to various flux-tubes at different longitudes and
latitudes due to the tangled, variable magnetic structures on the star
\citep{cra08}.  On the other hand, the phase shift may also be caused
by a combination of differential rotation and slow migration of the
activity belt during a stellar dynamo cycle \cite{lan10}.

Note that we did not derive the ephemerides from our observational
data, but used available 2002 ephemerides provided by \cite{shk03}.
\cite{shk08} reported phase shifts at an amount of $- 0.17$ between
the 2003 and 2006 observations; therefore, (almost) the same amount of
shift may be applicable to our observations obtained in April 2006
at the McDonald Observatory.  Figure~10 shows the results by including
the $-0.17$ phase shift, which produces an appropriate correlation
between our limited number of observational data and the best-fit
bright-spot models.  The peak point of the planet induced activity is
estimated by \cite{shk03} to occur when the planet is about at the
third quadrature ($\phi = 0.8$).  This primary estimation was based
on best-fit bright-spot models over first available data points.
However, the lack of a sufficient number of observational data and
the detected phase shifts created a large uncertainty about the phase
of the peak point.

The latest four ephemerides of HD~179949 system are listed in Table~3.
Calculated phases from the four different epochs are listed in Table~4.
Table~5 shows the mean phase shifts (local time), noting that averages
were attained if there was more than one observation per night.
The inconsistency of calculated phases from different epochs may imply
a change in mechanical properties in the system as function of time due
to star-planet interaction.  Figure~11 shows the change in phase as
a function of time, where it shows a low-order polynomial trend.
Data points are computed from
${\Delta}{\phi} = \phi - \phi_{\rm Butler}$, where $\phi$ are the
phases given in Table~4 and $\phi_{\rm Butler}$ represent the data
of \cite{but06}.  The trend shown in Fig.~11 may be crucial for
understanding the nature of the HD~179949 star-planet system;
however, further discussions on this subject should be deferred
until additional studies and observations have become available.
However, it is noteworthy that the existence of similar phase
shifts have previously been identified in other star-planet systems,
notably $\tau$~Boo and HD~189733 \citep{shk08}, possibly indicative
of stellar magnetic dynamo evolution \citep{lan10}.


\begin{figure}
\centering
\begin{tabular}{c}
\epsfig{file=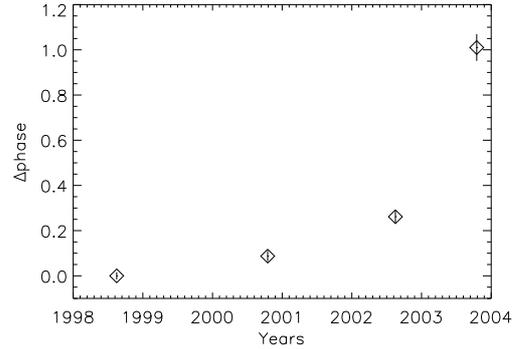,width=0.95\linewidth}
\end{tabular}
\caption{
Phase shift difference relative to the observations
obtained in July 1998.
}
\end{figure}

\begin{table}[h]
\begin{center}
\caption{Mean Phase Shifts (Local Time)$^a$\label{tab5}}
\begin{tabular}{lcccc}
\noalign{\smallskip}
\hline
\hline
\noalign{\smallskip}
Date & $\phi (1)$ & $\phi (2)$ & $\phi (3)$ & $\phi (4)$ \\
\noalign{\smallskip}
\hline
\noalign{\smallskip}
   April 5, 2006  &   0.2676	&  0.0163  &  0.1906  &  0.2776 \\
   April 6, 2006	&   0.5878	&  0.3365  &  0.5108  &  0.5978 \\
   April 7, 2006	&   0.9141	&  0.6628  &  0.8370  &  0.9241 \\
   April 8, 2006	&   0.2362	&  0.9850  &  0.1591  &  0.2463 \\
   April 9, 2006	&   0.5647	&  0.3135  &  0.4877  &  0.5749 \\
  April 11, 2006	&   0.2078	&  0.9566  &  0.1307  &  0.2180 \\
\noalign{\smallskip}
\hline
\end{tabular}
\medskip\\
$^a$$\phi (1)$, $\phi (2)$, $\phi (3)$, and $\phi (4)$ correspond to the observations obtained on
07/1998, 09/2000, 07/2002, and 09/2003, respectively, (see Table~3) with $\phi (1) \equiv
\phi_{\rm Butler}$.
\end{center}
\end{table}


\section{Conclusions}

The aim of our study was to extend the observational database of
planet induced stellar emission (PIE) due to the interaction of close-in
exoplanets and their host stars by focusing on HD~179949.  We obtained
twelve observations during April 2006 at the 2.7~m telescope of the
McDonald Observatory covering three planetary orbits.
The PIE phenomenon previously suggested by \cite{cun00} was first observed
in HD~179949 \citep{shk03} concerning Ca~II H and K.  It is now known to
be present in five other star-planet systems; see \cite{cun02} for an early
comparison between theory and observations.  Besides Ca~II H and K, the PIE
effect has also been identified in other spectral features including the
Ca~II infrared triplet \citep{saa01} and in coronal X-rays
\citep{kas08,saa08}.

Our observations appear to confirm the activity enhancement for the Ca~II K
core between phases 0.4 and 0.9.  \cite{shk08} reported the disappearance
of the PIE phenomenon during their 2003 and 2006 observation runs.  However,
we have an insufficient number of data to be able to confirm or refute this
finding.  Nevertheless, the two relevant data points obtained in our
April 2006 observing run show the PIE effect to be present in the
Ca~II K core consistent with the previous results by Shkolnik and
collaborators.  Additionally, we obtained several data points
showing the absence of star-planet interaction consistent with the
previous predictions.  This latter finding is consistent with the
interpretation that the PIE effect is due to magnetic star-planet
interaction presumably facilitated along the Parker spiral of the
stellar wind \cite[e.g.][]{saa04}.  Recent magnetospheric simulations
for the interaction between stellar and planetary magnetic fields and
outflows were given by \cite{coh11} taking HD~189733 as an example,
which point to the formation of associated Ca II line signatures,
expected to occur in a stochastic manner.

Regarding our observations, the Ca~II H core analysis did not reveal
any significant fluctuation, which is at odds with the prediction based on our Ca~II~K detection.  Also, HD~179949
was observed at low altitude, hence at a relatively great air mass,
during the observation run in April 2006, which is not suitable for
detecting relatively weak activity enhancements.  Previously,
\cite{shk03} deduced the Ca~II H core fluctuation to be about
2/3 of the intensity of the fluctuation observed in the K core.
The Al~I $\lambda$3944~{\AA} line indicated no notable
fluctuations during the time increments when the Ca~II K emission
enhancements occurred.  This is strong evidence that the observed
emission is indeed due to the close-in giant planet, as expected.
However, there is an exceptional Al~I data point obtained on 
April 11 possibly revealing significant emission in the stellar photosphere
without a Ca~II K chromospheric counterpart.  Thus, the cause for
this event is apparently inherent in the star, unrelated to the
existence of the planet.

\section*{Acknowledgments}
This work was supported in part by NASA through the 
American Astronomical Society's Small Research Grant Program.
Additional support was provided by the Theodore Dunham, Jr.
Grant of the Fund for Astrophysical Research.  This paper includes
data taken at The McDonald Observatory of The University of Texas
at Austin.


\end{document}